\documentclass[smallextended]{svjour3}       % onecolumn (second format)
\smartqed  % flush right qed marks, e.g. at end of proof
\usepackage{graphicx}
\usepackage[cmex10]{amsmath}
\usepackage{amsmath}
\usepackage{amssymb,colortbl}

\begin{document}

\title{Satellite Laser Ranging and General Relativity measurements in the field of the Earth: state of the art and perspectives%\thanks{Grants or other notes
}


\author{ 
	David Lucchesi$^{1,2,3}$         \and
        Luciano Anselmo$^{2}$\and 
        Massimo Bassan$^{3,4}$\and
        Carmelo Magnafico$^{1,3}$\and
        Carmen Pardini$^{2}$\and
        Roberto Peron$^{1,3}$ \and
        Giuseppe Pucacco$^{3,4}$ \and
        Ruggero Stanga$^{5}$ \and
        Massimo Visco$^{1,3}$%etc.
}


\institute{David Lucchesi \at
\email{david.lucchesi@iaps.inaf.it}  \\
 Tel.: +0039-06-49934385 
\\
\at ${}^{1}$ Istituto Nazionale di Astrofisica (INAF) - Istituto di Astrofisica e Planetologia Spaziali (IAPS), Via del Fosso del Cavaliere, 100, 00133 Roma, Italy     \\
\at ${}^{2}$ Istituto di Scienza e Tecnologie della Informazione (ISTI) - Consiglio Nazionale delle Ricerche (CNR), Via Moruzzi, 1, 56124 Pisa, Italy \\
\at ${}^{3}$ Istituto Nazionale di Fisica Nucleare (INFN), Sezione di Roma Tor Vergata, Via della Ricerca Scientifica 1, 00133 Roma, Italy\\
\at ${}^{4}$ Dipartimento di  Fisica, Universit\`a di Roma Tor Vergata,   00133 Roma, Italy\\   
\at ${}^{5}$ Istituto Nazionale di Fisica Nucleare (INFN), Sezione di Firenze, Via Giovanni Sansone 1, 50019 Sesto Fiorentino, Firenze, Italy
}

\date{Received: date / Accepted: date}

\maketitle

\begin{abstract}
Recent results of the LARASE research program in terms of models improvements and relativistic measurements are presented. In particular, the results regarding the development of new models for the non-gravitational perturbations that affect the orbit of the LAGEOS and LARES satellites are described and discussed. These are subtle and complex effects that need a deep knowledge of the structure of the satellite and of its physical characteristics in order to be correctly accounted for. In the field of gravitational measurements, we introduce a new and precise measurement of the relativistic Lense-Thirring precession. The role of the errors related to the knowledge of the gravitational field of the Earth in this kind of measurements is also discussed. Finally, the main results in relativistic measurements and constraints obtained during the last few years by means of the laser tracking of passive satellites are summarized. The key role of the Satellite Laser Ranging technique in these activities is highlighted, together with the sinergy it produces between space geodesy and fundamental physics measurements.

\end{abstract}

\section{Introduction}
\label{intro}

Laser ranging to Cube Corner Retroreflectors (CCRs) on Earth's orbiting satellites as well as to the CCRs arrays placed on the Moon by the Apollo and Luna missions has provided a number of impressive results in the study of both Earth and Moon geophysics, i.e. on Earth and Lunar sciences, as well as in gravitational physics and in Einstein's General Relativity (GR). These powerful laser tracking techniques are known, respectively, as Satellite Laser Ranging (SLR) and Lunar Laser Ranging (LLR)  \cite{2002AdSpR..30..135P,1973Sci...182..229B,1994Sci...265..482D,Merkowitz2010}. The observable is represented by the round-trip time of short laser pulses measured by means of very precise time devices down to a resolution of about 10 ps or less in the case of the SLR, and about 100 ps in the case of the LLR (1 ps = \(1 \times 10^{-12}\)~s). The root-mean-square (RMS) of the corresponding range measurements is down to a mm precision for SLR and about 10 mm precision for LLR.

It is worth of mention that both SLR and LLR are two very important techniques, among the others, of the observational space geodesy. Indeed, Very Long Baseline Interferometry (VLBI), Global Navigation Satellite Systems (GNSS), Doppler Orbitography and Radiopositioning Integrated by Satellite (DORIS), together with SLR and LLR constitute the Global Geodetic Observing System (GGOS)~\cite{2010JGeo...49..112R}. Obviously, in the context of the GGOS, improvements in technology and in modeling will produce advances in geodesy and geophysics as well as in GR measurements. Therefore, these two important research fields, space geodesy and GR, are not independent, but tightly related, and improvements in one field produce benefits in the other, and vice versa, in a positive loop. For instance, in the near future it is expected an improvement of one order of magnitude in the accuracies in the observational components of space geodesy \cite{2012ivs..conf..357C}. This implies a tenfold advancement in a series of closely related fields, such as: modeling, measuring techniques and their accuracy, reference frames and their realization, ground stations geometry. Consequently, from these advances, the final accuracy of fundamental physics measurements using the main space geodesy techniques will also benefit.

In this paper we deal with SLR. In particular, we focus on the main issues regarding the GR measurements to be performed by laser tracking to geodetic and passive satellites \cite{1996NCimA.109..575C,1997EL.....39..359C,1997CQGra..14.2701C,1998Sci...279.2100C,2004Natur.431..958C,2010PhRvL.105w1103L,2014PhRvD..89h2002L,2016EPJC...76..120C,2017mas..conf..131L}. These concern the models and the estimate of the systematic sources of error in view of new measurements and tests of gravitational physics that can be achieved in the future with this powerful technique \cite{Lucchesietal2015}. 

SLR is currently limited by the satellite center-of-mass offset, by the atmosphere refraction and by stations biases. Improvements in these fields will allow a sub-mm precision in the RMS of the SLR range measurements with very significant benefits in all geophysical and GR measurements. 

Of course, a fundamental point within this context is played by the dynamical models used for the reconstruction of the satellites orbit. Therefore, all the activities that aim to improve the models presently developed for both gravitational and non-gravitational perturbations --- as well as those activities with the goal to develop new models in this direction --- are welcome and play a significant role. In particular, the non-gravitational perturbations modeling is an important and delicate task, since their effects are very subtle and complex to model in a reliable manner. Moreover, a step forward along the issues described above is an important prerequisite for the construction of an estimate of the systematic errors --- in the case of new measurements of relativistic effects --- that be reliable, robust and accurate.

The activities of LARASE (LAser RAnged Satellites Experiment) experiment are here described and discussed. The final goal of LARASE is to verify the predictions of GR in its weak-field and slow-motion limit with respect to those of alternative theories of gravitation \cite{1993tegp.book.....W,2014grav.book.....P,2014LRR....17....4W}. To reach this important objective by means of precise and accurate measurements, LARASE has started a deep review of the non-gravitational perturbations (NGP) that act on the orbit of the two LAGEOS and LARES satellites, as well as an analysis of the various models to be used for the main gravitational perturbations.

The rest of the paper is organized as follows. In Section \ref{sec:ngp_mod}, we describe the main activities performed to improve the dynamical models used to handle the subtle effects produced by the NGP. We focus on the spin model of the satellites, the drag from the neutral atmosphere and the Yarkovsky-Schach thermal effect. In Section \ref{sec:pod}, our recent results for the precise orbit determination (POD) of the considered satellites are shown. In Section \ref{sec:LT}, a new measurement of the Lense-Thirring effect is presented. In Section \ref{sec:gra_mod}, we present some recent results regarding the evaluation of the errors for the measurement of the Lense-Thirring effect related to the background gravitational field. In Section \ref{sec:rel}, the state of the art of the relativistic measurements performed so far by means of laser-ranged satellites is summarized. Finally, in Section \ref{sec:con} our conclusions and recommendations on the subjects discussed are provided.

\section{Models for the non-gravitational perturbations}
\label{sec:ngp_mod}

In this Section we introduce and discuss some of the improvements we achieved in the development of new models for the main NGP that perturb the orbits of the two LAGEOS~\cite{1985JGR....90.9215C} satellites as well as that of LARES~\cite{2013AcAau..91..313P}. We focus on the knowledge of the internal structure of the satellites, in particular on their moments of inertia, on the spin dynamics and, finally, on the perturbations due to neutral drag and Yarkovsky-Schach effect.

\subsection{Internal structure}
\label{sec:ngp_mod-int}

Starting from the original technical documentation of the LAGEOS and LARES satellites, we have been able to reconstruct information about their structure, used materials, and moments of inertia \cite{2016AdSpR..57.1928V}. One important goal of this work has been to have an independent estimate of the moments of inertia of the satellites that, it has to be noticed, were not measured on the flight models of the two LAGEOS satellites. The degree of deviation from the spherical symmetry of a satellite is a very important quantity to be measured (or at least calculated), since it causes a gravitational torque, similar to the Hipparchus precession of the Earth, that contributes to the spin evolution of the satellite (see Section \ref{sec:spin}). The analysis reported in \cite{2016AdSpR..57.1928V} is also very important for the development of a refined thermal model in order to properly account for the complex perturbation produced by the thermal thrust effects (see Section \ref{sec:ther}) which, in turn, depend on the spin evolution.

In Table \ref{tab:mom} the moments of inertia of the two LAGEOS satellites, as computed in \cite{2016AdSpR..57.1928V} with normalized densities, are shown. As we can see, the two satellites have almost the same oblateness, about 0.04.

%%%%%%%%%%%%%%%%%%%%%%%%%%%%%%%%
\begin{table}[!h]
\caption{\label{tab:mom}Principal moments of inertia of LAGEOS and LAGEOS II in their flight arrangement.}
\centering
\begin{tabular}{lccc}
\hline
  Satellite &\multicolumn{3}{c} {Moments of inertia  (kg~m\(^2\))} \\
	& \(I_{xx}\) & \(I_{yy}\) & \(I_{zz}\) \\
\hline\\
  LAGEOS  & $11.42\pm0.03$ &$10.96\pm0.03$&$10.96\pm0.03$  \\
		\hline
	LAGEOS  II  & $11.45\pm0.03$&$11.00\pm0.03$&$11.00\pm0.03$ \\
	\hline
\end{tabular}
\end{table}
%%%%%%%%%%%%%%%%%%%%%%%%%%%%%%%%

\begin{figure*}[t!]
\centering
\includegraphics[width=8cm]{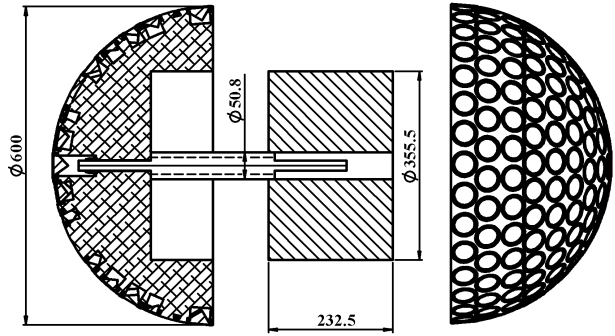}
\caption{ The LAGEOS satellites assembly. The dimensions are in mm. The two aluminum hemispheres are shown with the section of the cavities containing the CCRs together with the internal brass cylinder and the copper-beryllium shaft.}
\label{sezione-3a}
\end{figure*}

A section view of the two satellites is shown in Figure \ref{sezione-3a}, where the main parts of the structure are visible: i) two hemispheres of aluminum containing the CCRs, ii) the brass core, which contributes to increase the mass of the satellite, iii) the copper-beryllium shaft that fastens the different parts of the satellites.

%%%%%%%%%%%%%%%%%%%%%%%%%%%%%%%%%%%%%%%%%%%%%%%%%%%%%%%%%%%%%%%%
%%%%%%%%%%%%%%%%%%%%%%%%%%%%%%%%%%%%%%%%%%%%%%%%%%%%%%%%%%%%%%%%

This work has been very useful also to understand on one side, and to correct on the other side, some contradictory information provided in the historical literature of the older LAGEOS about some of its characteristics, such as the internal dimensions, the material of the inner cylinder and of the shaft. We refer to \cite{2016AdSpR..57.1928V} for further details. In the case of LARES, adopting for its mass the official value of 386 kg and assuming a spherical mass distribution, we obtained a moment of inertia of \(4.77 \pm0.03\) kg~m\(^2\). Anyway, an oblateness as small as 0.002 is possible.

\subsection{Spin dynamics}
\label{sec:spin}
The spin of the satellites plays a major role in the modeling of thermal effects, since these depend on the satellite attitude with respect to the radiation sources (see Section \ref{sec:ther}).  Models of the spin evolution of the LAGEOS satellites are found in the literature since 1991 \cite{1991JGR....96.2431B,1996JGR...10117861F,2004JGRB..109.6403A},  but they work successfully  only in the 
fast-spin approximation:  they provide solutions based on averaged equations, that are valid as long as the satellite rotation period is much smaller  than the orbital one. \\
We developed  a new complete model, named LASSOS (LArase Satellites Spin mOdel Solutions), to describe the evolution of the spin of the LAGEOS and LARES satellites \cite{2018arXiv180109098V}.  The LASSOS model provides a solution for the spin evolution valid for any value of the satellite rotational period. In the fast-spin limit, LASSOS correctly reproduces the results of previous models.

\begin{figure*}[th!]
\centering
\includegraphics[width=8cm]{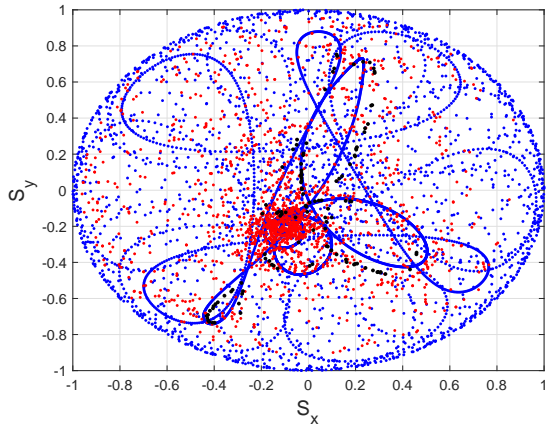}
\caption{Spin evolution of LAGEOS II on the basis of the LASSOS model (blue dots), compared with observations (red and black).
 Projection of the satellite spin direction on the equatorial plane  (\(S_x,S_y\)) in the J2000 reference frame.}
\label{Spin_SxSy}
\end{figure*}

\begin{figure*}[h!]
\centering
\includegraphics[width=8cm]{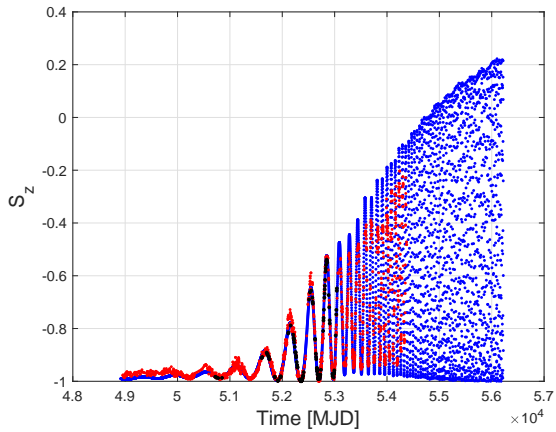}
\caption{Spin evolution of LAGEOS II on the basis of the LASSOS model (blue dots). Component of the satellite spin direction along the vertical axis (\(S_z\)) of the J2000 reference frame as a function of time expressed in Modified Julian Date (MJD).}
\label{Spin_Sz}
\end{figure*}

\begin{figure*}[t!]
\centering
\includegraphics[width=8cm]{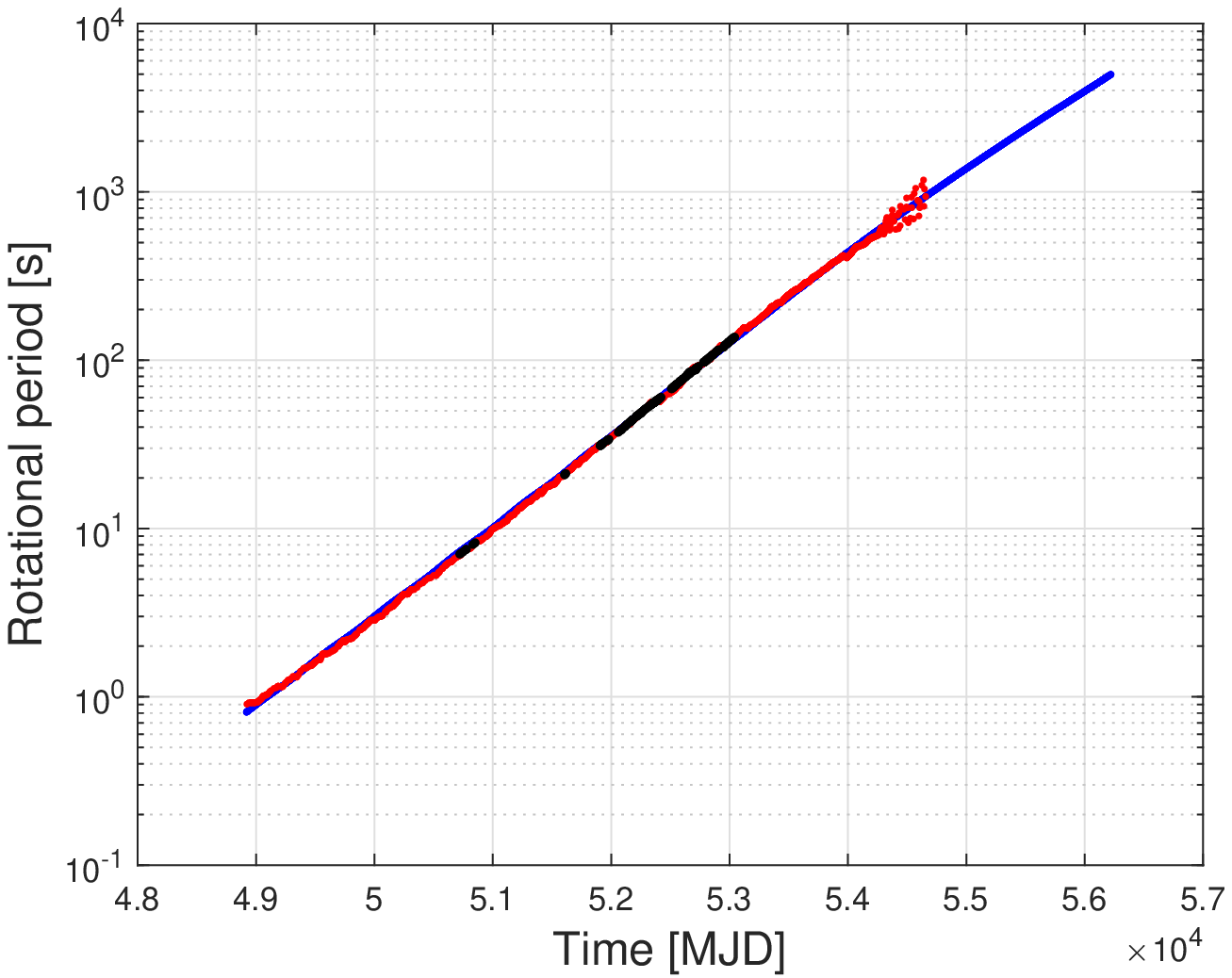}
\caption{Spin evolution of LAGEOS II on the basis of the LASSOS model (blue dots). Rotational period (\(P\)) of the satellite as a function of time expressed in Modified Julian Date (MJD).}
\label{Spin_periodo}
\end{figure*}

%We developed  a new complete model, named LASSOS (LArase Satellites Spin mOdel Solutions), to describe the evolution of the spin of the LAGEOS and LARES satellites \cite{2018arXiv180109098V}. The spin model plays a major role in the modeling of thermal effects, since they depend on the satellite attitude with respect to the radiation sources (see Section \ref{sec:ther}). 
% 
%The main new feature of our model is its ability to completely reproduce the spin evolution of LAGEOS and LARES satellites in the general case, with no restriction to a fast-spin approximation.  In particular, we solved the full set of  Euler equations of motion considering not-averaged torques. Past solutions for the spin model of the LAGEOS satellites were successful only using averaged equations \cite{1991JGR....96.2431B,1996JGR...10117861F,2004JGRB..109.6403A}.
%In fact, in the fast-spin approximation the solutions provided by the models based on averaged equations are valid as long as the satellite rotation period is much smaller  than the orbital one. Conversely, the LASSOS model provides a solution for the spin evolution valid for any value of the satellite rotational period.
%
In the case of LAGEOS II, in Figures \ref{Spin_SxSy}, \ref{Spin_Sz} and \ref{Spin_periodo} we show, respectively, the two equatorial components (\(S_x,S_y\)) of the spin direction of the satellite at different times, the time evolution of the projection \(S_z\) of the spin direction along the vertical axis, and the time evolution of the satellite rotational period \(P\). The reference frame is the J2000 frame. 
In these figures we compare the predictions of the model (blue dots) with the available observations (in red those spectrally determined from the SLR data \cite{2013AdSpR..52.1332K}, and in black the collection of previous observations as reported in \cite{2007Andres}).

%%%%%%%%%%%%%%%%%%%%%%%%%%%%%%%%%%%%%%%%%%%%%%%%%%%%%%%%%%%%%%%%
%%%%%%%%%%%%%%%%%%%%%%%%%%%%%%%%%%%%%%%%%%%%%%%%%%%%%%%%%%%%%%%%

%%%%%%%%%%%%%%%%%%%%%%%%%%%%%%%%%%%%%%%%%%%%%%%%%%%%%%%%%%%%%%%%
%%%%%%%%%%%%%%%%%%%%%%%%%%%%%%%%%%%%%%%%%%%%%%%%%%%%%%%%%%%%%%%%

%%%%%%%%%%%%%%%%%%%%%%%%%%%%%%%%%%%%%%%%%%%%%%%%%%%%%%%%%%%%%%%%
%%%%%%%%%%%%%%%%%%%%%%%%%%%%%%%%%%%%%%%%%%%%%%%%%%%%%%%%%%%%%%%%

As we can see, over the time span where the observations of spin orientation and rate are available, the agreement between our model and the experimental data is quite good. We refer to \cite{2018arXiv180109098V} for further details on the LASSOS model as well as for the results regarding LAGEOS and LARES.

\subsection{Neutral drag}
\label{sec:drag}

Since the orbit of LARES is much lower than that of the two LAGEOS, with an height of about 1450 km vs 5900 km, it was expected a much stronger impact of the drag from the neutral atmosphere on the orbit of this satellite \cite{Lucchesietal2015,7180629}. This has motivated us to start an accurate analysis of the effects of the neutral atmosphere on the orbit of LARES. In particular, we started from an evaluation of the decay of the semi-major axis of the satellite over a given time span.

In this activity, we jointly used the software GEODYN II \cite{1998pavlis} together with a modified version of SATRAP (SATellite Reentry Analysis Program) \cite{Pardini1994,2012P&SS...67..130P}. The main feature of SATRAP is the use of several different models for the Earth's atmosphere together with the appropriate geomagnetic and solar activity indices. In this way we have been able to investigate directly the impact of the neutral drag on the orbit of LARES using some of the best models for the Earth's atmosphere that are available in the literature \cite{2017mas..conf..131L,PARDINI2017469}. 

By means of GEODYN we performed a POD of LARES over a time span of about 3.7 years (from April 6, 2012 to December 25, 2015). In the set of dynamical models we did not include, beside the neutral drag, the thermal effects. We measured an orbital decay  of the satellite semi-major axis of about 2.74 mm/d (i.e. close to 1 m/yr). Such decay corresponds to a transversal (mean) acceleration of about \(-1.444\times 10^{-11}\) m/s\(^2\). We then included the neutral drag in the dynamical model and, by means of a least-squares fit of the tracking data, we estimated the drag coefficient \(C_\mathrm{D}\) of the satellite. We obtained a mean value of about 4 for the drag coefficient.

With SATRAP we then performed an additional analysis over the same time span with the goal of estimating the drag coefficient independently from the value estimated with GEODYN. In this analysis we accounted for the measured decay --- by assuming the unmodeled transversal acceleration that was estimated with GEODYN as reference --- and considered several atmospheric models. We obtained \(C_\mathrm{D} \lesssim 4\), in very good agreement with the value  estimated with GEODYN, see \cite{PARDINI2017469}. 

This means that the current best models developed to account for the behaviour of the neutral atmosphere  are able to reproduce the observed decay of LARES, within their intrinsic errors (around 15\%) and their range of applicability. Among the atmospheric models, we considered the Jacchia-Roberts 1971 model (JR-71) \cite{1976STIN...7624291C}, the Mass Spectrometer and Incoherent Scatter radar 1986 model (MSIS-86) \cite{1987JGR....92.4649H}, the Mass Spectrometer and Incoherent Scatter radar Extended 1990 model (MSISE-90) \cite{1991JGR....96.1159H}, the NRLMSISE-00 model \cite{2002JGRA..107.1468P}, developed at the US Naval Research Laboratory (NRL) and the GOST-2004 model \cite{Volkov2004}, issued by the State Committee on Standardization and Metrology of the Russian Federation. Two of these models, MSIS-86 and JR-71, are also implemented in GEODYN.

We remark that, after modeling the neutral drag in GEODYN, a very small decay was still present in the integrated residuals of the LARES semi-major axis, corresponding to a residual, unaccounted for,  transversal acceleration of about \(-2\times 10^{-13}\) m/s\(^2\). We refer to \cite{PARDINI2017469} for further details.

Consequently, unlike what happened in the past for the two LAGEOS satellites, where the Earth-Yarkovsky thermal drag and charged particle drag \cite{rub82,1990JGR....95.4881R} were the leading causes of the observed decay (\(\approx 90\%\)), with our analysis we found that the drag from the neutral atmosphere is able to explain \(\approx 99\%\) of the observed decay of the LARES semi-major axis. This means that a new analysis must be performed to extract from the observed decay a possible smaller contribution related to other unmodeled effects, such as the thermal ones, acting on the satellite. In this context, a new analysis becomes necessary to determine the contribution of the drag and of the thermal effects to the residuals of the orbital elements of LARES.

\begin{figure*}[t!]
\centering
\includegraphics[width=8cm]{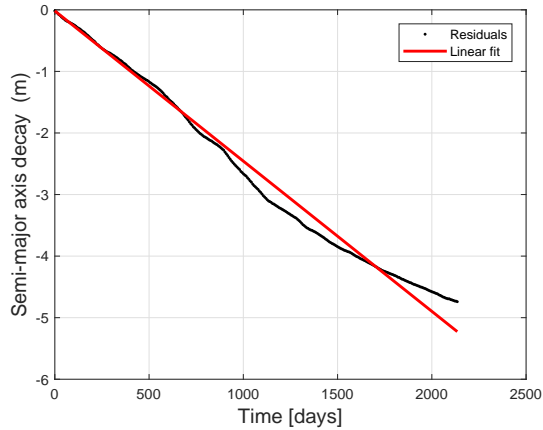}
\caption{Semi-major axis decay of LARES over a time span of 2135 days (about 5.8 years). The black dots represent the integrated residuals of the satellite semi-major axis. The residuals \cite{2006P&SS...54..581L} have been determined by means of a POD obtained with GEODYN II. The dynamical model of GEODYN neither includes the neutral and charged atmosphere drag, nor the thermal effects. The red line represents a linear best fit to the integrated residuals used to evaluate a decay of about 2.44 mm/d.}
\label{LR-decay}
\end{figure*}

We have recently undertaken this study recently, and extended also to the two LAGEOS satellites. In Figure \ref{LR-decay}, a new measurement for the decay of LARES semi-major axis is shown over a period of about 5.8 years, from April 13, 2012 (MJD 56030). The measured decay is about 2.44 mm/d and corresponds to an average transversal acceleration of about \(-1.289\times 10^{-11}\) m/s\(^2\).

%%%%%%%%%%%%%%%%%%%%%%%%%%%%%%%%%%%%%%%%%%%%%%%%%%%%%%%%%%%%%%%%
%%%%%%%%%%%%%%%%%%%%%%%%%%%%%%%%%%%%%%%%%%%%%%%%%%%%%%%%%%%%%%%%

\begin{figure*}[h!]
\centering
\includegraphics[width=8cm]{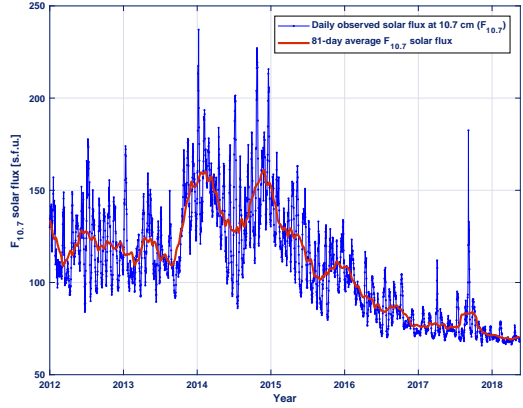}
\caption{Solar activity since the launch of LARES. Daily observed solar flux at 10.7 cm and its 81-day values averaged over three solar rotations: from January 1, 2012 to May 15, 2018.}
\label{solar}
\end{figure*}

The smaller value obtained for the semi-major axis decay over the longer time span of this analysis is not a surprise. In fact, while the first analysis of about 3.7 years was performed during a maximum of the Sun activity, the extended time span is characterized by a deep decrease in solar activity. Indeed, solar activity is approaching a very deep minimum, probably the deepest since these measurements are performed, see Figure \ref{solar}\footnote{The daily values of the observed solar flux can be found at the following address  \(https://www.ngdc.noaa.gov/stp/space-weather/solar-data/solar-features/solar-radio/noontime-flux/penticton/penticton_observed/tables/\) of the NOAA's National Centers for Environmental Information.}. This indirectly confirms, once more, how precise these measurements, based on SLR data, are.

%%%%%%%%%%%%%%%%%%%%%%%%%%%%%%%%%%%%%%%%%%%%%%%%%%%%%%%%%%%%%%%%
%%%%%%%%%%%%%%%%%%%%%%%%%%%%%%%%%%%%%%%%%%%%%%%%%%%%%%%%%%%%%%%%

%We therefore reviewed the drag effects on the orbit of these satellites and we started a number of  different activities as: i) the comparison of the predictions of the different atmospheric models at the altitudes of the satellites, ii) the estimate of the perturbing accelerations acting on the satellites, iii) the estimate of the disturbing effects on their orbit, and iv) the estimates of their drag coefficient \(C_D\). We focus here on the last point, i.e., on the decay of the semi-major axis of LARES.

\subsection{Thermal effects}
\label{sec:ther}

An intricate role, among the NGPs, is played by the subtle thermal effects that arise from the radiation emitted from the satellite surface as consequence of the non-uniform distribution of its temperature. In the literature of the older LAGEOS satellite this problem was considered since the early 80s’ of the past century to explain the (apparently) anomalous behaviour of the along-track acceleration of the satellite, characterized by a complex pattern \cite{1987JGR....9211662R,1987JGR....92.1287R,1988JGR....9313805R,1989AnGeo...7..501A,1990JGR....95.4881R,1990A&A...234..546F,1991JGR....96..729S,1996CeMDA..66..131S,1996P&SS...44.1551F,1997JGR...102..585R,1997JGR...102.2711M,1999AdSpR..23..721M,2002P&SS...50.1067L,2007Andres}. The dynamical problem to solve is quite complex and should take into account the following main aspects:
\begin{itemize}
\item a deep physical characterization of the various elements that constitute the satellite, i.e., a very good knowledge of the emission and absorption coefficients, of the thermal conductivity, of the heat capacity, of the thermal inertia, \(\dots\);
\item a very good knowledge of the rotational dynamics of the satellite, i.e., of its spin orientation and rate;
\item a reliable model for the radiation sources: Sun and (especially) Earth.
\end{itemize}
We have tackled the problem following the two approaches considered in the past (but with some differences): we first developed a simplified thermal model based on averaged equations, as in \cite{1988JGR....9313805R,1989AnGeo...7..501A,1990A&A...234..546F,1996P&SS...44.1551F} then we developed a general thermal model not restricted to averaged equations, as in \cite{1996CeMDA..66..131S,2007Andres}.

In the first (simplified) case the thermal model is mainly based on: i) the application of the energy balance equation on the satellite surface, and ii) a linear approximation for the distribution of the temperature over the satellite surface (restricted to its CCRs) with respect to its equilibrium (mean) temperature.

Conversely, in the second (more general) case the thermal model is mainly based on: i) a satellite (metallic structure) in thermal equilibrium, ii)
 the CCRs rings are at the same temperature of the satellite metallic body, and iii) for each CCR the thermal exchange with the satellite is computed. 
 
In the case of the two LAGEOS, in agreement with the description of Section \ref{sec:ngp_mod-int}, for the metallic structure we considered both the external aluminum surface and the internal brass core plus the copper-beryllium shaft. For both models, simplified and general, it is mandatory to have a reliable model for the evolution of the spin vector of the satellite.

The two main thermal perturbations to be taken into account are the solar Yarkovsky-Schach \cite{rub82,1990A&A...234..546F,1991JGR....96..729S,1996P&SS...44.1551F,1996CeMDA..66..131S} effect and the Earth-Yarkovsky (or Rubincam) effect \cite{rub82,1987JGR....92.1287R,1990JGR....95.4881R}.

The first effect is characterized by an anisotropic emission of thermal radiation that arises from the temperature gradients across the surface that are produced by the solar heating and the thermal inertia of the various elements of the satellite (mainly the CCRs). This perturbation is responsible for long-term effects on the orbit of a satellite when the thermal radiation is modulated by the eclipses. 

In the case of the second effect, the temperature gradients responsible for the anisotropic emission of thermal radiation are produced by the Earth's infrared radiation. The bulk of the effect is due to the CCRs and their thermal inertia. Also this perturbation is responsible for long-term effects on the orbital elements of a satellite.

In the following, we provide our results for LAGEOS II in the case of the simplified thermal model applied to the solar Yarkovsky-Schach thermal effect. The analysis has been performed on a time span of 4080 days starting from October 25, 1992 (MJD 48920). The LASSOS model has been used for the spin model evolution of the satellite, see Section \ref{sec:spin}.

\begin{figure*}[h!]
\centering
\includegraphics[width=12cm]{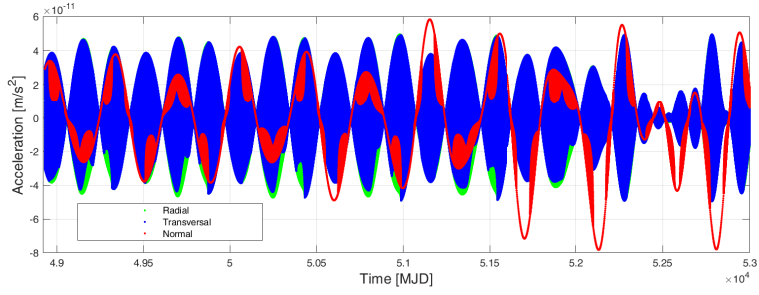}
\caption{Perturbing accelerations on LAGEOS II due to the Yarkovsky-Schach effect over a time span of about 11.2 years. These results have been obtained by our simplified thermal model.}
\label{L2-acc}
\end{figure*}

In Figure \ref{L2-acc} are shown the perturbing accelerations in the Gauss reference frame, i.e. the radial (green), transversal (blue) and normal, or out-of-plane (red) components of the Yarkovsky-Schach acceleration.
These results are in very good agreement with those obtained by \cite{2007Andres} and based on a more general model for the thermal effects.

%%%%%%%%%%%%%%%%%%%%%%%%%%%%%%%%%%%%%%%%%%%%%%%%%%%%%%%%%%%%%%%%
%%%%%%%%%%%%%%%%%%%%%%%%%%%%%%%%%%%%%%%%%%%%%%%%%%%%%%%%%%%%%%%%

\begin{figure*}[h!]
\centering
\includegraphics[width=12cm]{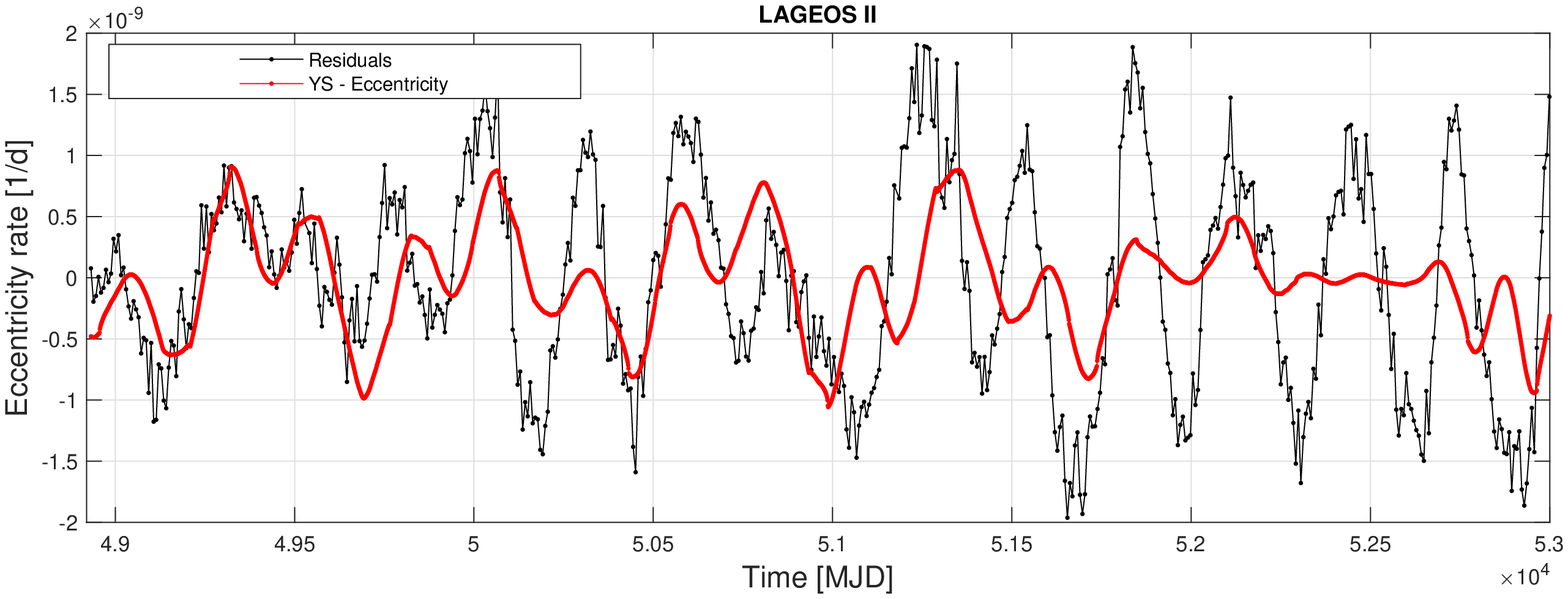}
\caption{Yarkovsky-Schach effect on LAGEOS II eccentricity. The predictions of the simplified thermal model (red line) are compared with the residuals in the eccentricity (black line) obtained by means of a POD performed over the same time span of 4080 days.}
\label{L2-ecc}
\end{figure*}

\begin{figure*}[t!]
\centering
\includegraphics[width=12cm]{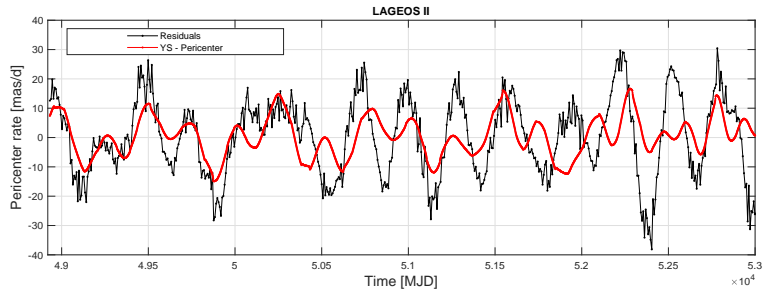}
\caption{Yarkovsky-Schach effect on LAGEOS II argument of pericenter. The predictions of the simplified thermal model (red line) are compared with the residuals in the argument of pericenter (black line) obtained by means of a POD performed over the same time span of 4080 days.}
\label{L2-peri}
\end{figure*}

In Figures \ref{L2-ecc} and \ref{L2-peri}, the perturbing effects on the satellite eccentricity and argument of pericenter are shown and compared with the residuals in these elements obtained by an independent POD performed by us.

%%%%%%%%%%%%%%%%%%%%%%%%%%%%%%%%%%%%%%%%%%%%%%%%%%%%%%%%%%%%%%%%
%%%%%%%%%%%%%%%%%%%%%%%%%%%%%%%%%%%%%%%%%%%%%%%%%%%%%%%%%%%%%%%%

%%%%%%%%%%%%%%%%%%%%%%%%%%%%%%%%%%%%%%%%%%%%%%%%%%%%%%%%%%%%%%%%
%%%%%%%%%%%%%%%%%%%%%%%%%%%%%%%%%%%%%%%%%%%%%%%%%%%%%%%%%%%%%%%%

In this analysis, the EIGEN-GRACE02S solution has been used as model of the gravitational field of the Earth~\cite{2005JGeo...39....1R}. The arc length was 7 days, no empirical accelerations\footnote{Empirical accelerations are kinematic terms introduced to absorb unmodeled effects.} were used and the thermal effects (Yarkovsky-Schach and Earth-Yarkovsky) were not modeled. General relativity was modeled with the exception of the Lense-Thirring effect. More details on the analysis strategy can be found in Section~\ref{sec:pod}.

As we can see, the agreement with the residuals is good, but not perfect. In fact, the agreement could not be perfect since, besides the signature of the Yarkovsky-Schach effect, these residuals also contain the signature of the Earth-Yarkovsky effect as well as that of the asymmetric reflectivity of the satellite's hemispheres~\cite{2003GeoRL..30.1957L,2004CeMDA..88..269L}.

In a forthcoming paper, the details of both thermal models developed by LARASE, the simplified one and the general one, will be provided in the case of the two LAGEOS satellites and compared with independent analyses performed with GEODYN.

\section{Precise orbit determination}
\label{sec:pod}

The powerful SLR technique provides normal points for the two LAGEOS and LARES satellites with a RMS precision at the mm level. Consequently, a comparable precision is needed for the models included in GEODYN and for the models used \textit{a posteriori} for the analysis of the orbital residuals, as in the case of the thermal thrust forces, see Section~\ref{sec:ther}.

A reliable POD of the two LAGEOS and LARES satellites represents a very important goal and an essential prerequisite within the LARASE activities. Outcome of this procedure is a precise orbit of the satellite obtained by fitting the tracking data with a suitable set of dynamical models, along with series of estimated parameters. The orbit so determined is the starting point to compute the residuals in the Keplerian elements~\cite{2006P&SS...54..581L} of the satellites and then build with them the physical observables that contain the imprint of the unmodeled effects of the Einstein's theory of GR \cite{1996NCimA.109.1709C}. The computation strategy employed by us (which is customary in the space geodesy community) foresees the split of the entire analysis time period in smaller intervals, called arcs; for each arc, a POD is performed and the residuals in the Keplerian elements computed. It is important to underline that, in this strategy,the relativistic part (indeed, only the Lense-Thirring part in the analysis here described) need not be included in the set of dynamical models, and the corresponding effect (if any) on the orbital elements is measured analyzing the residuals time series.

The setup we are currently using is the one shown in Table~\ref{tab:modelli}. It accounts for: i) the satellite dynamics, ii) the measurement procedure and iii) the reference frames transformations. In this context, our strategy is to align our models, wherever possible, with the international resolutions and conventions, such as the International Astronomical Union (IAU) 2000 Resolutions~\cite{2003AJ....126.2687S} and the IERS Conventions (2010)~\cite{2010IERS-Conv-2010}.

%%%%%%%%%%%%%%%%%%%%%%%%%%%%%%%%%%%%%%%%%%%%%%%%%%%
\begin{table*}[h!]
\caption{Models currently used by the LARASE research program for the analysis of the orbit of the two LAGEOS and LARES satellites. The models are grouped in gravitational perturbations, non-gravitational perturbations and reference frames realizations.}
\label{tab:modelli}
\centering
\begin{tabular}{lll}
\hline\noalign{\smallskip}
Model for & Model type & Reference \\
\noalign{\smallskip}\hline\noalign{\smallskip}
Geopotential (static) & EIGEN-GRACE02S/GGM05S & \cite{2005JGeo...39....1R,Tapley2013,JGRB:JGRB50058} \\
Geopotential (time-varying, tides) & Ray GOT99.2 & \cite{1999Ray} \\
Geopotential (time-varying, non tidal) & IERS Conventions (2010) & \cite{2010IERS-Conv-2010} \\
Third--body & JPL DE-403 & \cite{1995Standish} \\
Relativistic corrections & Parameterized post-Newtonian & \cite{2003AJ....126.2687S,1990CeMDA..48..167H} \\
\noalign{\smallskip}\hline\noalign{\smallskip}
Direct solar radiation pressure & Cannonball & \cite{1998pavlis} \\
Earth albedo & Knocke-Rubincam & \cite{1987JGR....9211662R} \\
Earth-Yarkovsky & Rubincam (1987-1990) & \cite{1987JGR....92.1287R,1988JGR....9313805R,1990JGR....95.4881R} \\
Neutral drag & JR-71/MSIS-86 & \cite{1976STIN...7624291C,1987JGR....92.4649H} \\
Spin & LASSOS (2015-2017) & \cite{2018arXiv180109098V} \\
\noalign{\smallskip}\hline\noalign{\smallskip}
Stations position & ITRF2008 & \cite{2011JGeod..85..457A} \\
Ocean loading & Schernek and GOT99.2 tides & \cite{1998pavlis,1999Ray} \\
Earth Rotation Parameters & IERS EOP C04 & \cite{IERS-EOP_C04} \\
Precession & IAU 2000 & \cite{2003A&A...412..567C} \\
Nutation & IAU 2000 & \cite{2002JGRB..107.2068M} \\
\noalign{\smallskip}\hline
%\multicolumn{3}{l}{\textsuperscript{\(\ast\)}We emphasize that selected parts of these post-Newtonian corrections have not been included in the modellization setup used for specific analyses of relativistic effects.} \\
%\multicolumn{3}{l}{\textsuperscript{\(\ast\)}Obviously, the post-Newtonian corrections have not been included among the models used}\\
%\multicolumn{3}{l}{ within specific analyses of relativistic effects measurements.} \\
\end{tabular}
\end{table*}
%%%%%%%%%%%%%%%%%%%%%%%%%%%%%%%%%%%%%%%%%%%%%%%%%%%
%
In Table~\ref{T1}, we provide our estimate of the mean orbital elements for LARES and the two LAGEOS satellites, calculated from the satellites ephemerides obtained from a POD.

%%%%%%%%%%%%%%%%%%%%%%%%%%%%%%%%%%%%%%
\begin{table*}[h!]
  \caption[]{Mean orbital elements of LAGEOS, LAGEOS II and LARES.}
  \centering
  \begin{tabular}{@{}cccrrr@{}}
  \hline
     
$\rm{Element}$   & Unit   & Simbol 						&\rm{LAGEOS}     & \rm{LAGEOS II} & \rm{LARES} \\
\hline
 semi-major axis \, & [{\rm km}] & {\it a}                  			& 12 270.00          & 12 162.07 		&   7 820.31     \\
 eccentricity  & & {\it e}                                 				&  0.004433          & 0.013798       	&   0.001196 \\
 inclination \, & [{\rm deg}] & \rm{i}                           			&  109.84              & 52.66       		& 69.49  \\
\hline
\end{tabular}
\label{T1}
\end{table*}
%%%%%%%%%%%%%%%%%%%%%%%%%%%%%%%%%%%%%%

\begin{figure*}[h!]
\centering
\includegraphics[width=12cm]{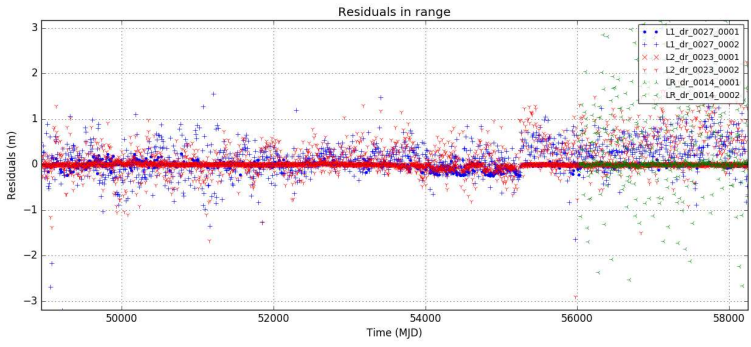}
\caption{Range residuals of LAGEOS (blue), LAGEOS II (red) and LARES (green).}
\label{residui}
\end{figure*}

\begin{figure*}[h!]
\centering
\includegraphics[width=12cm]{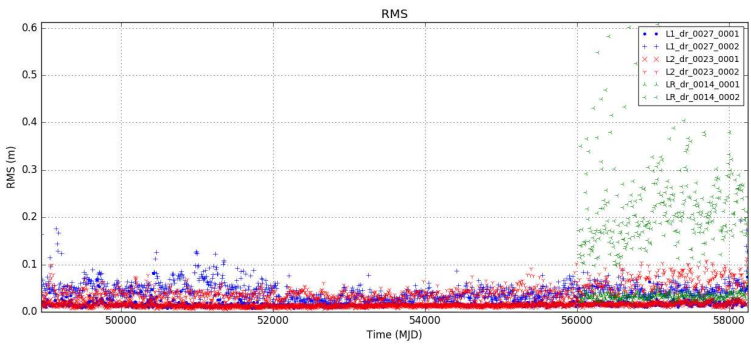}
\caption{RMS of the range residuals of LAGEOS (blue), LAGEOS II (red) and LARES (green).}
\label{rms}
\end{figure*}

In figures~\ref{residui} and~\ref{rms} we show the results of the POD for the considered satellites in two cases: i) when empirical accelerations have been estimated for each arc (label 0001), and ii) when empirical accelerations have not been included in the data reduction (label 0002). These are two of the standard runs that are routinely performed by the LARASE collaboration\footnote{In these particular analyses, we use the EIGEN-GRACE02S model for the gravitational field of the Earth, an arc length of 7 days and all tracking stations have been weighted equally. Usually, the radiation coefficient \(C_\mathrm{R}\) of the satellites is estimated with the empirical accelerations.}. Neither the Lense-Thirring effect nor the thermal effects have been modeled. With the colors blue, red and green are shown, respectively, the results for LAGEOS (L1), LAGEOS II (L2) and LARES (LR). The two analyses cover a time span longer than 25 years in the case of the two LAGEOS satellites, from October 30, 1992 (MJD 48925) to February 16, 2018 (MJD 58165).

%%%%%%%%%%%%%%%%%%%%%%%%%%%%%%%%%%%%%%%%%%%%%%%%%%%%%%%%%%%%%%%%
%%%%%%%%%%%%%%%%%%%%%%%%%%%%%%%%%%%%%%%%%%%%%%%%%%%%%%%%%%%%%%%%

%%%%%%%%%%%%%%%%%%%%%%%%%%%%%%%%%%%%%%%%%%%%%%%%%%%%%%%%%%%%%%%%
%%%%%%%%%%%%%%%%%%%%%%%%%%%%%%%%%%%%%%%%%%%%%%%%%%%%%%%%%%%%%%%%

The first plot shows the range residuals (computed as mean values over each arc) of the three satellites, while the second plot shows their RMS. Of course, in the case of the second run (with no empirical accelerations), the scatter of the residuals is larger, and so is their RMS. In Table~\ref{pod}, the statistics of the two different analyses for the three satellites in terms of mean (M) and standard deviation ($\sigma$) of the range residuals and of their RMS is given. In the case of the two LAGEOS, the statistics refers to the entire period of the analysis, i.e. to about 25 years.

%%%%%%%%%%%%%%%%%%%%%%%%%%%%%%%%%%%%%%
\begin{table*}[h!]
  \caption[]{POD statistics over the time span of the analyses. The mean (M) and the standard deviation ($\sigma$) of the range residuals and of their RMS are provided for the two analyses performed with and without the use of empirical accelerations (EA). The unit is cm.}
  \centering
  \begin{tabular}{@{}ccccccccc@{}}
  \hline
 \multicolumn{1}{c}{} &\multicolumn{4}{c}{ With EA (0001)} & \multicolumn{4}{c}{Without EA (0002)	} \\
% \hline
\multicolumn{1}{c}{} &  \multicolumn{2}{c}{Residuals} & \multicolumn{2}{c}{RMS} &  \multicolumn{2}{c}{Residuals} & \multicolumn{2}{c}{RMS} \\
\hline
   & M & $\sigma$ & M & $\sigma$  & 	M	 &  $\sigma$  & M &   $\sigma$     \\            
 LAGEOS  & \(-0.62\) & 5.98 & 2.29 & 1.52 & 15.00 & 65.69 & 5.77 & 3.56 \\   
 LAGEOS II  & \(-0.76\) & 3.53 & 1.48 & 0.40 & 21.17 & 45.76 & 4.43 & 1.72 \\ 
 LARES  & 0.14 & 4.06 & 3.34 & 0.66 & 87.76 & 146.70 & 22.65 & 9.66 \\            
\hline
\end{tabular}
\label{pod}
\end{table*}
%%%%%%%%%%%%%%%%%%%%%%%%%%%%%%%%%%%%%%

As we can see, when we use the empirical terms to absorb the unmodeled effects, we are able to fit the orbits of the satellites with RMS of few cm: about 1 or 2 cm for the two LAGEOS and \(\gtrsim 3\) cm for LARES.
However, it is important to underline that over shorter time spans --- and when all the GR effects are modeled --- the RMS level of the POD for the two LAGEOS satellites is usually better by a factor of two, in the range \(0.5-1\) cm.
We can see from Table \ref{pod} that the use of the empirical accelerations is effective in reducing the range residuals of the PODs to values compatible with a null average. 

{On the other hand, when no empirical terms are used, the lack of a good modeling is quite apparent, especially in the case of LARES, whose orbit is not determined better than a meter. 
 However, the use of empirical accelerations represents a workaround to handle unexplained phenomena that affect the POD; we believe that it is in the best interest of science (geodesy, geophysics, relativity etc.) to try to understand and model these phenomena, rather than fitting them away. }
 %At the same time, the use of empirical accelerations, especially in a rough and uncontrolled way, does not represent the best approach to improve the POD of a satellite and of the final products expected for space geodesy and geophysics, as well as for fundamental physics. 
Therefore, it is of the utmost importance to improve the dynamical model of the orbit of the satellites, especially for the non-gravitational forces, to reduce the use of the empirical accelerations by means of reliable models. 

\section{A new measurement of the Lense-Thirring effect}
\label{sec:LT}

The goals of LARASE in terms of relativistic measurements in the near-Earth environment are described in~\cite{Lucchesietal2015}. Among these measurements, a special attention is devoted to the orbit precession of a satellite produced by the so-called Lense-Thirring effect \cite{1918LenseThirring}. This orbit precession, namely of the right ascension of the ascending node \(\Omega\) and of the argument of pericenter \(\omega\), is produced by the Earth's angular momentum.

This effect is a consequence of the gravitomagnetic field of Einstein's GR, which takes part in generating \emph{spacetime} curvature together with the gravitoelectric field produced by  mass \cite{1973grav.book.....M}. Gravitomagnetism is a phenomenon analogous to (electro)magnetism, where mass currents play the same role of electric currents that generate the magnetic field. Therefore, mass currents in GR, together with mass, take part in determining the trajectory of a test body, i.e. its geodetic in \emph{spacetime}. It is worth to underline that gravitomagnetism plays a special role in the astrophysics of compact objects~\cite{Thorne1983} and it has (possible) cosmological consequences linked to Mach's Principle~\cite{1995grin.book.....C}. 

As for the two relativistic observables mentioned above, the right ascension of the ascending node is less perturbed by the long-term effects produced by the thermal thrust effects. For this orbital element the secular part of Lense-Thirring precession predicts:
\begin{equation}
\label{eq:Omega_sec_LT}
\dot{\Omega}^{\textrm{LT}} =  \frac{2GJ_{\oplus}}{c^2a^3(1-e^2)^{3/2}},
\end{equation}
where \(G\) represents the gravitational constant, \(J_{\oplus}\) the Earth's angular momentum and \(c\) the speed of light.
In Table~\ref{precessioni}, the value of the relativistic precession on the right ascension of the ascending node of each satellite is shown. These rates have been computed using the mean values of the orbital elements shown in Table~\ref{T1}.

%%%%%%%%%%%%%%%%%%%%%%%%%%%%%%%%%%%%%%%%%%%%%%%%%%%%%%%%%%%%%%%%%%%%%%%%%%%%%%%%%%%%%%%%%%%%%%%%%%%%%%%%%%%%%%%%%%%%%%%%%%
\begin{table}[h!]
  \caption{Rate (mas/yr) for the secular Lense-Thirring precession on the right ascension of the ascending node of the two LAGEOS and LARES satellites.}
  \centering
  \begin{tabular}{cccc}
  \hline
\rm{Orbital element}     	& \rm{LAGEOS}     	& \rm{LAGEOS II} 	& \rm{LARES} \\
\hline
%  \vspace{2mm}
\(\dot{\Omega}^{\textrm{LT}}\) & 30.67 & 31.50 & 118.48\\
\hline
\end{tabular}
\label{precessioni}
\end{table}
%%%%%%%%%%%%%%%%%%%%%%%%%%%%%%%%%%%%%%%%%%%%%%%%%%%%%%%%%%%%%%%%%%%%%%%%%%%%%%%%%%%%%%%%%%%%%%%%%%%%%%%%%%%%%%%%%%%%%%%%%%

Hereinafter, we focus on a new measurement of the Lense-Thirring effect that we recently performed on the combined analysis of the orbits of the two LAGEOS satellites and that of LARES. A POD of the orbit of the satellites was performed over a timespan of about 4.6 years, starting from April 6, 2012 (MJD 56023). For the arc length we used 7 days, no empirical accelerations were used and the thermal effects were not modeled. For the Earth gravitational field the GGM05S model~\cite{Tapley2013,JGRB:JGRB50058} was used. This is the model currently used by the ILRS community for their analyses.

{Following the procedure performed in the past in this context \cite{1996NCimA.109.1709C}, we  construct a linear combination of the three (node) observables that is insensible to the effects of two even zonal harmonics coefficients: these carry the largest uncertitude related  both to the static and (in part) to the dynamic part of the Earth's gravitational field (see Section~\ref{sec:gra_mod}).}
%combined the three (node) observables in such a way as to cancel the errors related both to the static and (in part) to the dynamic part of the Earth's gravitational field (see Section~\ref{sec:gra_mod}), namely of two even zonal harmonics coefficients, and extract from the adopted combination the relativistic precession for the combined orbits.

In the following equations, the mathematical expressions for the rate of the combined residuals, \(\delta \dot{\Omega}^{\textrm{res}}_{\textrm{comb}}\), and the GR prediction for the combined LT precession, \(\dot{\Omega}^{\textrm{LT}}_{\textrm{comb}}\), are explicitly given:
\begin{equation}
\label{eq:combi}
\delta \dot{\Omega}^{\textrm{res}}_{\textrm{comb}} \simeq \delta \dot \Omega_{\textrm{LI}}^\textrm{res} + k_1 \delta \dot \Omega_{\textrm{LII}}^\textrm{res} + k_2 \delta \dot \Omega_{\textrm{LR}}^\textrm{res} %+ \dots,
\end{equation}
\begin{equation}
\label{eq:combi_LT}
\dot{\Omega}^{\textrm{LT}}_{\textrm{comb}} =  \dot \Omega_{\textrm{LI}}^\textrm{LT} + k_1  \dot \Omega_{\textrm{LII}}^\textrm{LT} + k_2  \dot \Omega_{\textrm{LR}}^\textrm{LT} = 49.658 \quad\textrm{mas/yr}.
\end{equation}
The two coefficients \(k_1\) and \(k_2\) are obtained from the solution of a linear system of three equations in three unknowns in such a way to remove the errors from the first and third even zonal harmonics while solving for the relativistic precession.

However, 
%contrary to the approach used in previous measurements, based on the use of the satellites node as observable, 
{ unlike previous analyses based on the same observables} we combined these elements in such a way to cancel the errors related to the first, \(\bar{C}_{20}\), and third, \(\bar{C}_{60}\), even zonal harmonics (i.e. those with degree \(\ell\) equal to 2 and 6 and order \(m=0\)). Indeed, from the results of a number of preliminary analyses, we found that the estimation of the third even zonal harmonic of GGM05S is affected by larger uncertitudes than the second one\footnote{This holds at least on the time interval of the analysis we have performed.}, i.e. of the harmonic that describes the octupole deviation of the Earth's mass distribution from that of a perfect sphere. We refer to Section~\ref{sec:gra_mod} for further details.

Moreover, it is well known in the literature that the low-degree coefficients of the gravitational field of the Earth, and in particular the quadrupole coefficient\footnote{This is the coefficient that describes the main deviation of the Earth from the spherical symmetry.} \(\bar{C}_{20}\), are characterized by a significant time dependence, due to several phenomena that contribute to the variation of the moments of inertia of the Earth, with long-period effects of annual and inter-annual periodicity \cite{1997JGR...10222377C,2002Sci...297..831C,JGRB:JGRB50058,2018GeoJI.212.1218C}. 

For this reason, 
%in our POD, in the case of the quadrupole coefficient\footnote{This is the coefficient that describes the main deviation of the Earth from the spherical symmetry.} \(\bar{C}_{20}\), we did not follow the current recommendations of the IERS Conventions (2010)~\cite{2010IERS-Conv-2010}, but we used a value very close, in its time dependence, to the mean value we obtained from
{we used in our POD (differing from the current recommendations of the 2010 IERS Conventions ~\cite{2010IERS-Conv-2010}) a value for \(\bar{C}_{20}\)  that accounts for a time dependence as described by}
 a fit to the monthly solutions 
 %of this coefficient 
 provided by the Center for Space Research (CSR) of the University of Texas at Austin \cite{JGRB:JGRB50058,2018GeoJI.212.1218C}
\footnote{The CSR time series from GRACE RL05 are based on 15-day  estimates. These series also contain a forecast extension from the Joint Center for Earth Systems Technology (JCET) \cite{2009EGUGA..11.6584P,2009GSTMPavlis}.}. 
%See Section \ref{sec:gra_mod}.

\begin{figure*}[h!]
\centering
\includegraphics[width=10cm]{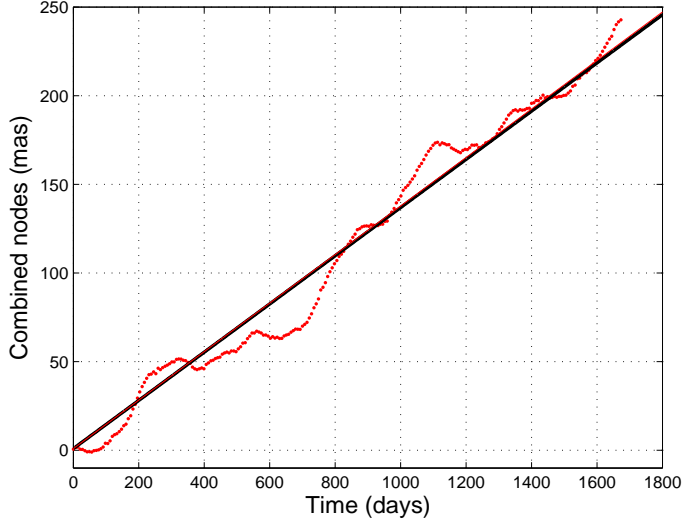}
\caption{Combined residuals (red dots) of the right ascension of the ascending node of LAGEOS, LAGEOS II and LARES.  The continuous red line represents a linear fit to the combined residuals. The continuous black line represents the corresponding prediction of GR for the precession.The starting epoch is April 6, 2012.}
\label{LT1}
\end{figure*}

The results for the integrated residuals of Eq. (\ref{eq:combi}) --- that we obtained by means of three independent POD of the satellites --- are shown in Figure~\ref{LT1}.

%%%%%%%%%%%%%%%%%%%%%%%%%%%%%%%%%%%%%%%%%%%%%%%%%%%%%%%%%%%%%%%%
%%%%%%%%%%%%%%%%%%%%%%%%%%%%%%%%%%%%%%%%%%%%%%%%%%%%%%%%%%%%%%%%

In this Figure, the integrated (and combined) residuals for the right ascension of the ascending node of the three satellites are shown over the time span of the analysis (red dots). The black line in the plot is not the result of a linear fit to the data, but it represents the GR prediction for the combined nodes, i.e. the integration in time of Eq. (\ref{eq:combi_LT}). Conversely, the red line represents the result of a linear fit to the residuals. As we can see, the fit is very close to the GR prediction and, at the scale of the plot, they almost overlap.

For the measured precession \(\dot{\Omega}_{\textrm{comb}}^{\textrm{meas}}\), which corresponds to the slope of the fitting line, we obtained a value of 49.966 mas/yr, very close to the GR prediction of Eq. (\ref{eq:combi_LT}). The fractional discrepancy with respect to the relativistic value is very small:
\begin{equation}
\label{eq:LT-frac}
\frac{\dot{\Omega}_{\textrm{comb}}^{\textrm{meas}}-\dot{\Omega}^{\textrm{LT}}_{\textrm{comb}}}{\dot{\Omega}^{\textrm{LT}}_{\textrm{comb}}} \simeq 6.2\times 10^{-3},
\end{equation}
indeed, it corresponds to a discrepancy of just 0.6\% with respect to GR prediction. Indeed, this result represents the most precise measurement achieved so far for the Lense-Thirring precession by means of a simple linear fit to the residuals of a POD\footnote{Obviously, the success of a simple linear fit is related to how many full cycles, of the unmodeled periodic effects, are contained in the time interval covered by the measurement.}.

{The combined residuals shown in Figure 12 are clearly characterized by the presence of unmodeled periodic effects. A spectral analysis of the residuals in the right ascension of the ascending node of the three satellites relates these periodic effects to the unmodeled thermal forces and to some tides. The main spectral lines are summarized in Table \ref{periodici}.}

%%%%%%%%%%%%%%%%%%%%%%%%%%%%%%%%%%%%%%%%%%%%%%%%%%%%%%%%%%%%%%%%%%%%%%%%%%%%%%%%%%%%%%%%%%%%%%%%%%%%%%%%%%%%%%%%%%%%%%%%%%
\begin{table}[h!]
  \caption{Periodic effects on the right ascension of the ascending node of the LAGEOS and LARES satellites. The periods have been rounded to the closest integer number and considered positive. The rate \(\dot{\lambda}\) refers to the time variation of ecliptic longitude of the Earth around the Sun.}
  \centering
  \begin{tabular}{cccc}
  \hline
\rm{Thermal effects}\textsuperscript{\(a\)} & \rm{LAGEOS} & \rm{LAGEOS II} & \rm{LARES} \\
\hline
\(\dot{\Omega}\) & 1052 & 570 & 211\\
\(2\dot{\Omega}\) &  526 & 285 & 105\\
\(\dot{\lambda}\) & 365 & 365 & 365 \\
\(2\dot{\lambda}\) &  183 & 183 & 183 \\
\(2(\dot{\Omega}-\dot{\lambda})\) & 280 & 111 & 67 \\
\(\dot{\Omega}+\dot{\lambda}\) &  271 & 953 & 497 \\
\hline
\rm{Solid tides} & \rm{LAGEOS} & \rm{LAGEOS II} & \rm{LARES} \\
\hline
165.565 & 911 & 622 & 217 \\
\hline
\rm{Ocean tides} & \rm{LAGEOS} & \rm{LAGEOS II} & \rm{LARES} \\
\hline
163.555 & 221 & 138 & 98 \\
\hline
\multicolumn{4}{l}{\textsuperscript{\(a\)}Some of these spectral lines are also common to solid and ocean tides.}\\
\end{tabular}
\label{periodici}
\end{table}
%%%%%%%%%%%%%%%%%%%%%%%%%%%%%%%%%%%%%%%%%%%%%%%%%%%%%%%%%%%%%%%%%%%%%%%%%%%%%%%%%%%%%%%%%%%%%%%%%%%%%%%%%%%%%%%%%%%%%%%%%%

By performing a non-linear fit to the residuals, including some of the unmodeled periodic effects, we have been able to further reduce the discrepancy with respect to the GR prediction.
We finally underline that this new measurement is less influenced by the systematic error sources with respect to previous measurements \cite{2016EPJC...76..120C,2017mas..conf..131L} (see Section \ref{sec:gra_mod}).
All these aspects will be further discussed and elaborated on in a forthcoming paper. 

\section{Gravitational perturbations and errors}
\label{sec:gra_mod}

It is well known from the literature that the main source of error in the case of relativistic measurements with passive laser-ranged satellites, as for the Lense-Thirring precession \cite{1996NCimA.109.1709C,1996NCimA.109..575C,2004GReGr..36.1321I,2006NewA...11..527C}, is given by the systematic effects related to the uncertainties of the even zonal coefficients \(\bar{C}_{\ell 0}\) of the expansion in spherical harmonics of the Earth's gravitational field.

In fact, the even zonal harmonics --- that measure the main deviation of the Earth's mass distribution from the spherical symmetry --- are responsible, just like the effect produced by the gravitomagnetic field, of a secular effect on the right ascension of the ascending node of a satellite \cite{1959AJ.....64..367K}:
\begin{equation}
\label{eq:pre_class}
\dot{\Omega}^{\textrm{class}} = -\frac{3}{2}n\left(\frac{R_{\oplus}}{a}\right)^2\frac{\cos i}{\left(1-e^2\right)^2} \left\lbrace -\sqrt{5}\bar{C}_{20}  +\dots \right\rbrace,
\end{equation}
where \(R_{\oplus}\) represents the Earth's mean equatorial radius, \(n\) the satellite mean motion and \(\bar{C}_{20}\) the normalized quadrupole coefficient\footnote{More precisely \(\dot{\Omega}^{\textrm{class}} =\sum_{\ell}\Omega_{\ell 0}\bar{C}_{\ell 0}\), where \(\Omega_{\ell 0}\) are the coefficients of the expansion.}. Consequently, the uncertainties in the knowledge of the even zonal harmonic coefficients --- as well as possible errors in their modeling because of their complex temporal evolution (especially for the very low degree coefficients \cite{1997JGR...10222377C,2002Sci...297..831C,JGRB:JGRB50058}) --- can alter or mask, or even mimic, the relativistic precession.

The trick of combining the relativistic observables~\cite{1996NCimA.109.1709C}, as shown in the previous Section by means of Eq. (\ref{eq:combi}), allows to reduce the impact of some of the errors related to both the static and dynamic part of the Earth gravitational field. {However, to provide a robust and reliable error budget for the relativistic measurement, it is illusory to apply the same combination to the errors associated to the remaining (and not cancelled) coefficients of the Earth gravitational field, even if these are calibrated and not simply formal errors}. Of course, the combination of Eq. (\ref{eq:combi}) should be used to estimate the errors, but through sensitivity analyses in dedicated PODs, and not by means of its crude application to the errors provided within a model for the gravitational field of the Earth.

\begin{figure*}[h!]
\centering
\includegraphics[width=10cm]{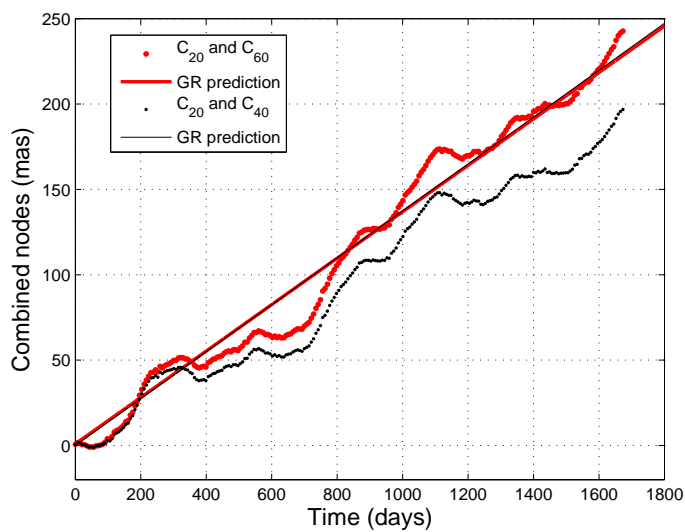}
\caption{Combined residuals (red and black dots) of the right ascension of the ascending node of LAGEOS, LAGEOS II and LARES. The two continuous red and black lines represent the corresponding predictions of GR for the Lense-Thirring precession: about 49.658 mas/yr for the first combination and about 50.175 mas/yr in the case of the second combination.}
\label{LT2}
\end{figure*}

{ As an example, we have computed the integrated residual of Eq.(\ref{eq:combi}) by removing, instead of the errors related to the \(\bar{C}_{20}\) and \(\bar{C}_{60}\)  coefficients, those related to \(\bar{C}_{20}\) and \(\bar{C}_{40}\), {as it was done in the past}.  This, obviously, defines a new set of equations and therefore two different parameters $k_1$ and $k_2$.
In Figure~\ref{LT2} we compare  these two results, {where the red dots refers to the combination that cancel the first and third harmonics, while the black dots refers to the combination that cancels the first and second harmonics}.}
%the integrated residuals of the satellites of Figure~\ref{LT1} --- i.e. those obtained after the removal of the errors related to the \(\bar{C}_{20}\) and \(\bar{C}_{60}\) coefficients (red dots) --- with those we obtain after the removal of the errors related to the \(\bar{C}_{20}\) and \(\bar{C}_{40}\) coefficients (black dots).

%%%%%%%%%%%%%%%%%%%%%%%%%%%%%%%%%%%%%%%%%%%%%%%%%%%%%%%%%%%%%%%%
%%%%%%%%%%%%%%%%%%%%%%%%%%%%%%%%%%%%%%%%%%%%%%%%%%%%%%%%%%%%%%%%

\begin{figure*}[h!]
\centering
\includegraphics[width=10cm]{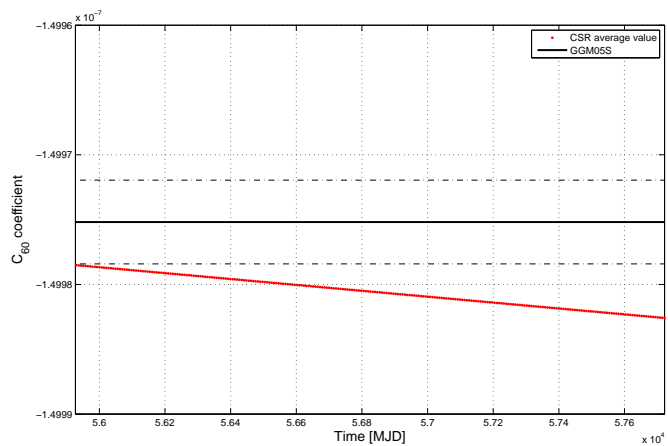}
\caption{Comparison between the mean evolution of the \(C_{60}\) coefficient as estimated by CSR (red dots), which follows a secular trend, with the static value provided in GGM05S (black line). The two dashed lines account for the calibrated error of the coefficient in GGM05S.}
\label{C60}
\end{figure*}

The two predictions of GR for the Lense-Thirring precession are very close (the red and black straight lines, 49.658 mas/yr vs 50.175 mas/yr), respectively when the third or second coefficient are removed together with the first, quadrupole one \footnote{{The difference arises from the different terms like Eq.(\ref{eq:pre_class}) that are removed from the solution.}}. However, the residuals obtained in the second case (removal of  \(\bar{C}_{40}\)) grow in time, unlike those obtained with the removal of the \(\bar{C}_{60}\) coefficient. 
This is in agreement with the fact, we underlined in the previous Section, that this even zonal harmonic is, at least on the time span of our analysis, worse in GGM05S with respect to the second one \(\bar{C}_{40}\), .
In fact, the time series for the third even zonal harmonic provided by the CSR \cite{2018GeoJI.212.1218C}, shows a significant discrepancy with respect to the static value given in GGM05S, see Figure \ref{C60}.

%%%%%%%%%%%%%%%%%%%%%%%%%%%%%%%%%%%%%%%%%%%%%%%%%%%%%%%%%%%%%%%%
%%%%%%%%%%%%%%%%%%%%%%%%%%%%%%%%%%%%%%%%%%%%%%%%%%%%%%%%%%%%%%%%

Conversely, if we apply Eq. (\ref{eq:combi}) to the calibrated errors of GGM05S up to the degree \(\ell=20\) (computed conservatively as sum of their absolute values) we obtain (for the static errors):
\begin{equation}
\label{eq:err_grav_zonali}
\varepsilon_{\textrm{sys}}^{\textrm{static}}\simeq 1.8\%,
\end{equation}
when the third harmonic \(\bar{C}_{60}\) is removed along with the first one, and
\begin{equation}
\label{eq:err_grav_zonali}
\varepsilon_{\textrm{sys}}^{\textrm{static}}\simeq 1.3\%,
\end{equation}
when the second harmonic \(\bar{C}_{40}\) is removed along with the first one. 
%These two estimates are clearly ambiguous and not reliable, since our preliminary analyses --- together with the estimates from the CSR --- show that it is the \(\bar{C}_{60}\) coefficient, and not the \(\bar{C}_{40}\), the one which is affected by a larger error in the GGM05S model when applied on the time interval of our measurement.
%
%For this reason, within the activities of LARASE we started a number of actions to provide a correct estimate for the various sources of systematic errors. In this context, our goal is to build an estimate of the errors 
{These error estimates, describing \(\bar{C}_{40}\) as affected by {a} larger {uncertainty} than \(\bar{C}_{60}\),  are in  contrast with our results  described above. This shows the need for a careful and thorough analysis of the various sources of systematic errors. And indeed, we have undertaken, within the activities of LARASE, a number of actions to provide a correct estimate of these systematics}
that will results in a robust and reliable error budget \cite{Lucchesietal2015,2017mas..conf..131L,LucchesiEGU2017p2}.

\section{Relativistic measurements with laser-ranged satellites}
\label{sec:rel}

In the field of relativistic measurements by means of the SLR technique and starting from the mid '1990s, gravitation, in the weak-field and slow-motion limit of GR, has been tested in one of its main aspects: the orbit precession of an Earth orbiting satellite produced by the gravitoelectric and gravitomagnetic fields of the planet. In particular, it was the latter and smaller field that was first measured~\cite{1996NCimA.109..575C}. Conversely, the relativistic (i.e. non-Newtonian) effects of the former field were measured many years later~\cite{2010PhRvL.105w1103L}.

These measurements, that involve two of the three Euler angles that define the orbit orientation with respect to an inertial frame, are known as Lense-Thirring precession and Schwarzschild (or Einstein) precession. Thanks to them, beside the cited relativistic precessions of GR, also a number of alternative theories for the gravitional interaction have been constrained in their predictions with respect to those of GR.

These alternative scenarios involve a possible violation of the inverse-square law~\cite{1986PhRvL..56....3F,2000PhRvD..61l2001N,2002PhRvD..66d6007D}, i.e. the existence of a new long-range interaction at the scale of the orbit of the satellites, but also the existence of non-symmetric or torsional contributions to the gravitational interaction~\cite{1979PhRvD..19.3554M,1988PhRvD..37..918M,2007PhRvD..76j4029M}.

In Table~\ref{LT-Sch} the best measurements obtained so far for both the Lense-Thirring and Schwarzschild precession are shown. In particular, the normalized precession for each measurement, \(\mu\) for the Lense-Thirring precession and \(\epsilon\) in the case of the Schwarzschild precession, is reported\footnote{Consequently, \(\mu\) and \(\epsilon\) equal to 1 means a perfect agreement with the predictions of GR.}.

%%%%%%%%%%%%%%%%%%%%%%%%%%%%%%%%%%%%%%
\begin{table*}[h!]
  \caption[]{Measurements of the Lense-Thirring secular precession due to the Earth gravitomagnetic field and of the Schwarzschild secular precession due to the Earth gravitoelectric field. 
    %Measurements of the Lense-Thirring and Schwarzschild secular precession due to the Earth gravitomagnetic and gravitoelectric fields. 
   The normalized Lense-Thirring precession \(\mu\) was obtained: combining the nodes of the two LAGEOS with the pericenter of LAGEOS II in the case of the measurement (1); combining the nodes of the two LAGEOS for the measurements (2) and (3); combining the nodes of the two LAGEOS with that of LARES for the other measurements. In the case of the Schwarzschild precession \(\epsilon\), the pericenter of LAGEOS II was used as the relativistic observable.}
  \centering
  \begin{tabular}{@{}ccccccccc@{}}
  \hline
\multicolumn{1}{c}{} & \multicolumn{4}{c}{ Lense-Thirring precession} & \multicolumn{4}{c}{Schwarzschild precession	} \\
% \hline
 \multicolumn{1}{c}{} & \multicolumn{1}{c}{\(\mu - 1\)} & \multicolumn{1}{c}{\(\delta\mu\)} & Year & Ref. &  \multicolumn{1}{c}{\(\epsilon - 1\)} & \multicolumn{1}{c}{\(\delta\epsilon\)} & Year & Ref. \\
\hline
 (1)&   \(+0.10\) & \(\pm 0.03\) & 1998 & \cite{1998Sci...279.2100C}  & 	\(+0.28\times 10^{-3}\)	 &  \(\pm 2.14\times 10^{-3}\)  & 2010 &   \cite{2010PhRvL.105w1103L}    \\            
 (2)& \(-0.01\) & \(\pm 0.02\) & 2004 & \cite{2004Natur.431..958C}  & \(-0.12\times 10^{-3}\) & \(\pm 2.10\times 10^{-3}\) & 2014 & \cite{2014PhRvD..89h2002L} \\   
 (3)& \(-0.01\) & \(\pm 0.01\) & 2004 & \cite{2004cosp...35..232L,2007AdSpR..39..324L} &  &  & &  \\ 
(4)& \(-0.006\) & \(\pm 0.002\)  & 2016 & \cite{2016EPJC...76..120C} &  &  &  &  \\ 
(5)& \(-0.001\) & \(\pm 0.001\)  & 2017 & \cite{2017mas..conf..131L} &  &  &  &  \\             
(6)& \(+0.006\) & \(\pm 0.010\)  & 2018 & This paper &  &  &  &  \\ 
\hline
\end{tabular}
\label{LT-Sch}
\end{table*}
%%%%%%%%%%%%%%%%%%%%%%%%%%%%%%%%%%%%%%
 
In the case of the Lense-Thirring effect measurements, the reported errors \(\delta \mu\) are  derived by the authors from a non-linear fit to the residuals and have to be considered as 1-\(\sigma\) formal errors. The only exceptions are the measurements (3) and (6), where a simple linear fit has been performed. Moreover, in the case of the present measurement (6), the error is conservatively provided by the bounds limits of the linear fit, and not by its formal error\footnote{As we specified in Section \ref{sec:LT}, a paper is in preparation with a more precise result based on a non-linear fit and with a detailed estimate of the systematic errors.}. All the above measurements have been also supplemented with an estimate of their systematic errors. Anyway, as discussed in Section~\ref{sec:gra_mod}, these estimates are not completely reliable, and were not included them in Table~\ref{LT-Sch}\footnote{The interested reader can refer to the cited literature for details.}.

In the case of the measurements of the Schwarzschild precession, the error \(\delta\epsilon\) is based on a sensitivity analysis of all the parameters involved in the non-linear fit performed, i.e. it is not a simple formal error. For these measurements, a detailed error budget of the systematic errors due to the main gravitational and non-gravitational disturbing effects has also been provided {\cite{2014PhRvD..89h2002L}}, at the level of \(2.5\%\).

In Table~\ref{tab:const}, the constraints obtained in~\cite{2014PhRvD..89h2002L} for possible alternative theories of the gravitational interaction are shown and compared with previous results. The physical parameters included in this Table are responsible, if they exist in nature, for a secular effect on the argument of pericenter of a satellite.  For the measurement of the advance of the pericenter of LAGEOS II, it was possible to carry on an accurate analysis of the systematic errors~\cite{2014PhRvD..89h2002L} : we can therefore provide (see Table~\ref{tab:const}), beside the errors obtained from a sensitivity analysis of the parameters of the fit, as previously described, also the error budget for each constraint: the last of the three values given for each constraint.

%%%%%%%%%%%%%%%%%%%%%%%%%%%%%%%%%%%%%%%
\begin{table*}[h!]%[H] add [H] placement to break table across pages
\caption{Summary of the constraints on fundamental physics obtained in \cite{2014PhRvD..89h2002L} with their corresponding errors. The constraints are compared with previous results~\cite{2005IJMPD..14.1657L,1992IJMPA...7..843C,2003PhLA..318..234L,2011PhRvD..83j4008M}.}
\label{tab:const}
%\begin{ruledtabular}
 \centering
  \begin{tabular}{@{}lcc@{}}
%\begin{tabular}{lcc{4.5cm}}
Parameter & Values or uncertainties & Previous values or uncertainties \\
\hline
\(|\alpha|\) & \(\simeq |(0.5 \pm 8.0) \pm 101|\times 10^{-12}\) & \(\pm 1\times 10^{-8}\) \\
\(\mathcal{C}_{\oplus \text{LII}}\) & \(\le (0.003 km)^4\pm (0.036 km)^4\pm(0.092 km)^4\) & \(\pm (0.16 km)^4\); \(\pm (0.087 km)^4\) \\
\(|2t_2 + t_3|\) & \(\simeq 3.5\times 10^{-4} \pm 6.2\times 10^{-3} \pm 7.49\times 10^{-2}\) & \(3\times 10^{-3}\) \\
\hline
\end{tabular}
%\end{ruledtabular}
\end{table*}
%%%%%%%%%%%%%%%%%%%%%%%%%%%%%%%%%%%%%%%
 
The parameter \(\alpha\) refers to the strength of a Yukawa--like potential predicted by some (possible) new long-range force\footnote{This is possible, in principle, if the metric tensor of GR is not the only tensor involved in the description of the gravitational interaction, but other fields (either scalar, vector, or tensor) are present~\cite{2002PhRvD..66d6007D}.}  for a characteristic range \(\lambda\) of the order of half the semi-major axis of LAGEOS II, i.e. close to the radius of the Earth. The parameter \(\mathcal{C}_{\oplus \text{LII}}\) refers to a possible additional interaction between the Earth and the satellite that comes to play in the case of a non-symmetric gravitation theory \cite{1979PhRvD..19.3554M}. If a non-vanishing torsional tensor is present, because of non-symmetric connection coefficients~\cite{1976RvMP...48..393H,2002RPPh...65..599H}, a  modification of  \textit{spacetime} will be produced\footnote{This represents a generalization of Einstein's GR when a Riemann-Cartan \textit{spacetime} is considered.}. The parameters \(t_2\) and \(t_3\) are two of a set of parameters that allow  to test for torsion and which appear in the metric~\cite{2007PhRvD..76j4029M}.

We finally remark that the measurement of the Schwarzschild precession, reported in Table~\ref{LT-Sch}, corresponds to a measurement of a combination of the post-Newtonian parameters \(\beta\) and \(\gamma\) \cite{1993tegp.book.....W}.

\section{Conclusions and recommendations}
\label{sec:con}

We have presented some of the results of the LARASE research program which are developed with the final goal to provide precise and accurate measurements of the gravitational interaction in the field of the Earth by means of laser measurements to passive satellites. In this framework, we have begun an activity aimed at improving the models of non-gravitational forces. 

We completed a new general model (named LASSOS) for the spin evolution of the two LAGEOS satellites and of LARES (Section~\ref{sec:spin}) and we are developing on a new and refined thermal model of the satellites to properly account  for the thermal thrust effects. These perturbing effects are mainly due to the Sun visible radiation (the Yarkovsky-Schach effect) and to the Earth's infrared radiation (the Earth-Yarkovsky effect) and critically depend on the spin evolution of each satellite.

We presented the results of a simplified thermal model, based on averaged equations: it provides  interesting results when its predictions for the evolution of the orbital elements of a satellite are compared with the results for the same elements independently obtained by a POD. In Section~\ref{sec:ther}, the results for LAGEOS II in the case of the solar Yarkovsky-Schach effect have been shown. A forthcoming paper will be devoted to a detailed and comprehensive description of both the simplified and the general thermal models.

Among the NGP, special attention was devoted to the effects of the atmospheric neutral drag. This perturbation is particularly important for LARES, because of its lower height with respect to the LAGEOS satellites. In Section~\ref{sec:drag}, our results for the decay of the semi-major axis of LARES have been shown. Also the relation between the decay and the varying solar activity has been higlighted. We have shown that almost all of the observed decay of the satellite semi-major axis can be explained by the effects of the neutral atmosphere. Anyway, it seems that a very small residual decay \(\lesssim 1\%\) is still unaccounted for. This may be due to the thermal effects and/or to the effects of the Coulombian interaction between the satellite potential and the charged particles in its surrounding environment. These issues are currently under investigation by LARASE.

We presented a preliminary result of our last measurement of the Lense-Thirring effect (Section \ref{sec:LT}). This new measurement is based on a simple linear fit to the integrated and combined residuals of the two LAGEOS and LARES satellites. A precise measurement, with a discrepancy of just 0.6\% with respect to the prediction of GR, was obtained.

Section~\ref{sec:gra_mod} was devoted to a discussion on the systematic sources of error related to the knowledge of the gravitational field of the Earth. This represents, for relativistic measurements with laser-ranged satellites, the main source of error and also one of the most complex to manage in a reliable way. Finally, in Section~\ref{sec:rel} the most important results obtained so far in terms of measurements and constraints of relativistic effects by means of laser-ranged satellites have been shown.

Thanks to the efforts of the ILRS, the precise measurements offered by the SLR technique cover a very important role for fundamental physics measurements, besides their ``\emph{institutional}'' role in space geodesy and geophysics. It will be very important that the quality and quantity of the laser observations are guaranteed and, possibly, improved in the future. 

This can be achieved by improving the quality of the ranging data, i.e. of the measurement of the round trip time of the laser pulses with a correct determination of the end points: on ground and on the satellite. This implies improvements in technology and in modeling, such as advancement in time counting devices and in calibration issues, in the knowledge of the stations  biases and of their position, as well as of the range correction and of the satellite as an ideal ``\textit{reflection point}''.

At the same time, it will be mandatory to increase the number of tracking stations in the future, since the number of satellites carrying CCRs is rapidly increasing, as the case of satellites used for navigation purposes clearly shows. Should the increase of the number of SLR stations, and of their quality, not be pursued, the result will be a reduction of the number of observations per satellite, with a consequent decrease in the quality of their orbit determination and of their final  products.

\begin{acknowledgements}
The authors acknowledge the ILRS for providing high quality laser ranging data of the two LAGEOS satellites and of LARES. Special thanks to the ILRS Analysis Centers and in particular to V.~Luceri and E.C.~Pavlis. This work has been in part supported by the Commissione Scientifica Nazionale II (CSNII) on astroparticle physics experiments of the Istituto Nazionale di Fisica Nucleare (INFN). 
\end{acknowledgements}

\bibliographystyle{spphys}       % APS-like style for physics
\bibliography{LARASE_JG}   % name your BibTeX data base

%\begin{thebibliography}{}
%\bibitem{Pearlman2002} Pearlman, MR, Degnan JJ, Bosworth JM, The International Laser Ranging Service, Adv Space Res \textbf{30} 135--43 (2002)
%
%\end{thebibliography}

\end{document}